\documentclass[prl,twocolumn,showpacs,floatfix]{revtex4}

\usepackage{graphicx}
\usepackage{array}
\usepackage{amsmath,amsfonts,amssymb}
\usepackage[ansinew]{inputenc}

\newcommand{\Figref}[1]{Fig.~\ref{#1}}

\begin{document}
\title{Narrow-line single-molecule transducer between electronic circuits and surface plasmons}


\author
{Michael C. Chong$^1$, Gaël Reecht$^1$, Herv\'e Bulou$^1$, Alex Boeglin$^1$, Fabrice Scheurer$^1$,\\
Fabrice Mathevet$^2$, Guillaume Schull$^{1\ast}$\\
\normalsize{$^1$Institut de Physique et Chimie des Mat\'eriaux de Strasbourg,} \\
\normalsize{UMR 7504 (CNRS -- Universit\'e de Strasbourg), 67034 Strasbourg, France} \\
\normalsize{$^2$Institut Parisien de Chimie Moléculaire, Chimie des Polymères,}\\
\normalsize{UMR 8232, CNRS - Université Pierre et Marie Curie, 94200 Ivry sur Seine, France} \\
\altaffiliation{guillaume.schull@ipcms.unistra.fr} 
}

\begin{abstract}

A molecular wire containing an emitting molecular center is controllably suspended between the plasmonic electrodes of a cryogenic scanning tunneling microscope. Passing current through this circuit generates an ultra narrow-line emission at an energy of $\approx$ 1.5 eV which is assigned to the fluorescence of the molecular center. Control over the linewidth is obtained by progressively detaching the emitting unit from the surface. The recorded spectra also reveal several vibronic peaks of low intensities that can be viewed as a fingerprint of the emitter. Surface-plasmon localized at the tip-sample interface are shown to play a major role on both excitation and emission of the molecular excitons. 
\end{abstract}

\date{\today}


\maketitle

Future ultrafast devices may rely on hybrid electronic-plasmonic circuitry \cite{Walters2010,Fedyanin2012}. A keystone for the realization of such components is the production of transducers allowing to controllably couple electrical and plasmonic signals \cite{Huang2014,Rai2013}. Tunneling electrons crossing the junction of a scanning tunneling microscope (STM) may act as an extremely localised excitation source for confined \cite{Berndt1991} or propagating \cite{Wang2011,Bharadwaj2011,Dong2015} surface plasmons. Tailoring the properties of this source appears as a dedicated method to controllably gate plasmonic waves \cite{Grosse2014} and to integrate nanoplasmonic architectures in electronic devices. Single-molecules may acts as controllable emitters capable of producing single-photons \cite{Brunel1999,Lounis2000} and narrow line emission \cite{Michaelis2000}, two mandatory characteristics for quantum computation applications. Qiu \textit{et al.} have demonstrated that a STM can be used to excite the luminescence of a single molecule \textit{decoupled} from the metallic substrate by a thin insulating layer and from the tip by vacuum \cite{Qiu2003}. Conserving these characteristics for a single-molecule integrated in a hybrid electro-plasmonic circuit is challenging because of the required \textit{direct} contact to metallic electrodes which alters the properties of the molecule and quenches its emission \cite{Avouris1984,Schneider2012,Reecht2014}. Recent works have shown that intrinsic radiative transitions may be preserved when elongated molecules are used as bridging elements \cite{Marquardt2010,Reecht2014}. In these examples, broad emission bands (FWHM $\geq$ 150 meV) are however observed, indicating the poor coherence of the emitted light.\\
Here we report on the narrow-line emission ($\approx$ 2.5 meV wide) from an electrically addressed molecular emitter connected to the plasmonic leads of a STM by short oligothiophene wires. Progressive lifting of the emitter from the substrate provides control over the linewidth of the emission which is finally limited by interactions with low energy phonons. Our fluorescence spectra also reveal vibronic features similar to Raman lines which provide a fingerprint of the molecular emitter. The detailed mechanisms that rule the complex interactions between the single-molecule emitter and localised surface plasmons is evidenced. Our results pave the way to nanoplasmonic devices integrating electrically gated molecular junctions as controllable excitation sources.\\

\begin{figure}
  \includegraphics[width=1.00\linewidth]{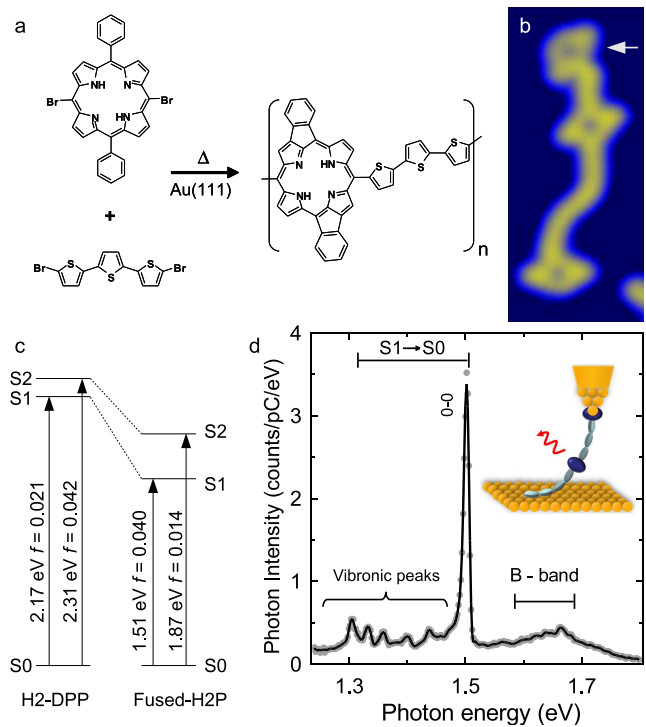}
  \caption{\label{fig1}\textbf{STM-induced luminescence of the suspended molecular wires.} (a) Sketch of the on-surface chemical reaction. (b) STM image (3.6 $\times$ 8.2 nm$^2$, $I$ = 0.1 nA, $V$ = -0.1 V) of a copolymer composed of oligothiophene and fused-H2P units polymerized on Au(111). (c) Optical gaps of H2-DPP and fused-H2P molecules calculated using a time-dependent density functional theory (TD-DFT) approach. The calculated oscillator strengths, $f$, of the respective transitions are also indicated (see details about the method in supplementary section S3 \cite{supp}). (d) Characteristic light emission spectra of a copolymer wire similar to the one in (b) suspended in the STM junction ($I$ = 4.2 nA, $V$ = 2 V, $z$ = 3.2 nm, acquisition time $t$ = 600 s, low resolution grating). The emission efficiency is limited to 5.10$^{-5}$ photon/electron and varies over orders of magnitude from junction to junction. The tip is brought into contact at the center of the fused-H2P marked by a white arrow. } 
\end{figure}

\noindent
We used a low temperature (4.5\,K) Omicron STM operating in ultrahigh vacuum for the experiment. The light collection setup is described in  [\citenum{Reecht2014}]. The detection setup is composed of a grating spectrograph (Princeton Instruments Acton Series SP-2300i) connected to a liquid nitrogen cooled CCD camera (Princeton instruments PyLoN-100BR-eXcelon). Two different gratings were used allowing for low (5 nm) and high (1 nm) spectral resolutions. The STM-tips were indented in the sample to cover them with gold and to modify their nanoscale shape in order to control their plasmonic response. 
The Au(111) samples were sputtered with argon ions and annealed. 5,5''-dibromo-2,2':5',2''-terthiophene and 5,15-(diphenyl)-10,20-(dibromo)porphyrin (\Figref{fig1}a) were successively deposited on the clean Au(111) surfaces at room temperature. As a next step, the samples were annealed at a temperature of $\approx$ 580 K in order to induce a polymerization reaction on the surface \cite{Grill2007,Lafferentz2012}.\\

\noindent  
STM images recorded after the copolymerization procedure suggest (see details in supplementary  section S1 \cite{supp}) that the formed nanowires (e.g., \Figref{fig1}b) are composed of a non-regular succession of covalently linked terthiophene and H2-DPP units. A more detailed analysis demonstrates (see supplementary section S1 \cite{supp}) that the porphyrin units also undergo an intramolecular cyclodehydrogenation \cite{Wiengarten2015} which leads to the formation of molecules (fused-H2P) where the two peripheral phenyl rings are fused to the porphyrin macrocycle (\Figref{fig1}a). Our time-dependent density functional theory (TD-DFT) simulations (\Figref{fig1}c) show that this cyclodehydrogenation reaction reduces the optical gap compared to the original H2-DPP.\\ 

\noindent
To decouple a fused-H2P from the metallic surface, the apex of the STM tip is brought into contact with the extremity of a copolymer strand (see supplementary section S2 \cite{supp}). The tip, with the attached wire, is then retracted from the surface by a distance large enough to suspend a single fused-H2P in the junction (inset of \Figref{fig1}d). Electrically-excited luminescence spectra obtained from such a molecular junction reveal a sharp and intense resonance at $h\nu = 1.51 \pm 0.04$ eV labelled 0-0 in \Figref{fig1}d. Together with the vibronic peaks that appear at lower energy, this feature can be assigned to the S1$\rightarrow$S0 transition of the fused-H2P molecule (\Figref{fig1}c). A less intense component (B-band) also appears at a higher energy $h\nu \approx 1.66$ eV. Here, the agreement with the calculated S0$\rightarrow$S2 transition (1.87 eV) is poor. Considering the oligothiophene side groups in our simulations reduces the S0$\rightarrow$S2 transition energy to 1.74 eV, leaving the S0$\rightarrow$S1 energy almost unchanged (see supplementary section S3 \cite{supp}). Alternatively, the B-band may be attributed to radiative transitions from excited vibrational states of S1 to S0 \cite{Wu2008,Dong2010}. Overall, the emerging picture is that the oligothiophene chains enable charge transport through the lifted copolymers while simultaneously ensuring the electronic decoupling of the central fused-H2P which acts as the optically active unit. 

\begin{figure}
  \includegraphics[width=1.00\linewidth]{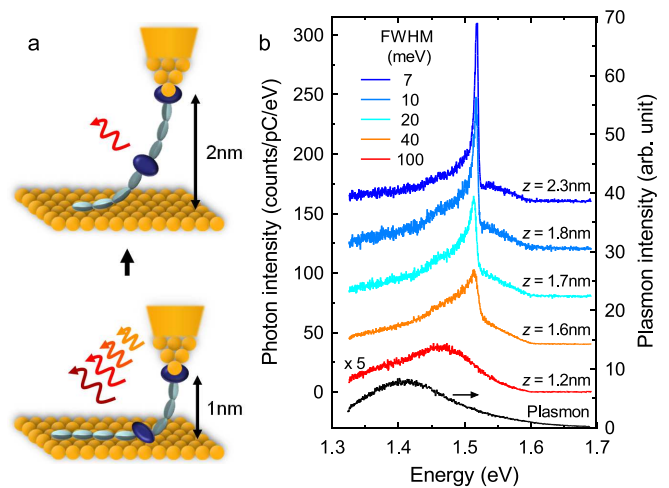}
  \caption{\label{fig2}\textbf{Controled emission linewidth} (a) Sketch of the experiment. (b) Light spectra as a function of the detachment of the molecular emitter from the surface ($V$ = 1.6 V, $t$ = 60 s, high resolution grating). The plasmon spectrum (black curve) is acquired with the same metal tip facing the bare Au(111) surface ($V$ = -3 V, $I$ = 9 nA, $t$ = 10 s).} 
\end{figure}   

\begin{figure}
  \includegraphics[width=1.00\linewidth]{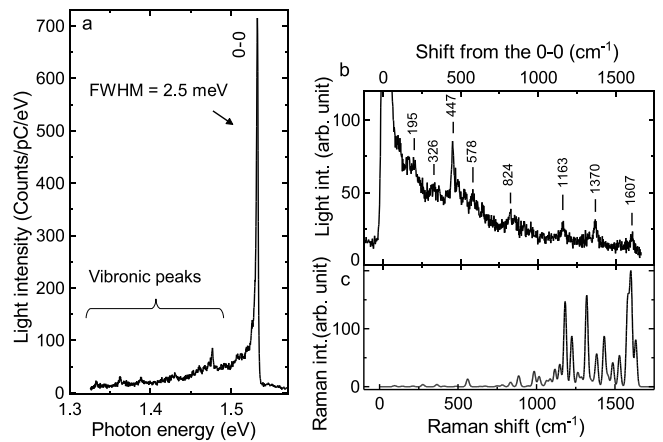}
  \caption{\label{fig3}\textbf{Vibronic spectra as spectroscopic fingerprint of the emitter} (a) Highly resolved light spectrum of a molecular junction ($V$ = 1.6 V, $I$ = 0.66 nA, $z$ = 1.8 nm, $t$ = 60 s, high resolution grating). (b) Same spectra where the light intensity is represented as a function of the shift from the 0-0 line. (c) Calculated Raman intensities for the vibrational modes of the fused-H2P (see details in supplementary section S4 \cite{supp}).} 
\end{figure}

\noindent 
We will now describe how the lifetime of the molecular excited state can be controlled by progressively detaching the emitting fused-H2P from the surface (\Figref{fig2}a). Figure \ref{fig2}b displays optical spectra of the S1$\rightarrow$S0 band of a suspended molecular wire, similar to the one in \Figref{fig1}a, for tip-sample distances ranging from $z$ = 1.2 nm to $z$ = 2.3 nm. For the shortest distance, the emitting fused-H2P is still adsorbed on the surface and the spectrum reveals a broad feature whose width ($\approx 100$ meV) is close to the one of the localised plasmon (black curve). At $z$ = 1.6 nm a sharper feature (FWHM = 40 meV) appears at the energy of the 0-0 transition. The next spectra show a progressive narrowing of this line that reaches a minimum width of 7 meV at $z$ = 2.3 nm, where the emitting fused-H2P is fully detached from the surface. The 0-0 line narrowing directly reflects an increase of the excited-state lifetime of the emitter, which is controlled by tuning the coupling between the molecular and the surface electronic states. As an extreme case, the spectrum of \Figref{fig3}a shows a spectral line $\approx$ 2.5 meV wide, to date the thinnest linewidth reported in a STM-induced light emission experiment. This corresponds to a lifetime of 0.26 ps and a coherence length of $\approx$ 0.15 mm, a value similar to the one of free-running laser diodes.   
Finally, the 0-0 emission line appears asymmetric in all spectra, revealing a sharp edge at high energy and a shallow tail at low energy. This spectral shape may be due to the coupling between the molecular exciton and low energy phonons of the molecular chains, a phenomenon that has been studied in detail for localised excitons in carbon nanotubes \cite{Galland2008}. This exciton-phonon coupling effectively limits the monochromaticity of the molecular emission.\\

\noindent
In previous STM induced-light emission experiments, vibronic features observed in optical spectra were associated to the harmonic progression of a single mode \cite{Qiu2003,Wu2008,Zhu2013}, and their relative intensities satisfied Franck and Condon factors. In contrast, our spectra reveal an extremely weak intensity of the vibronic peaks compared to the 0-0 contributions and a non linear dispersion of the states. This suggests that each line corresponds to a given vibrational mode of the emitter, similarly to what is observed in fluorescence line-narrowing spectroscopy of single molecules \cite{Fleury1995}. In Figure \ref{fig3}b the emitted intensity is represented as a function of the shift from the 0-0 line (in cm$^{-1}$) and tentitatively compared to (\Figref{fig3}c) the Raman intensities calculated at the DFT level for fused-H2P (see also supplementary sections S4 and S5 \cite{supp}). The high energy peaks are well reproduced by the theory and can be assigned to in-plane C-C stretching modes which do not exceed $\approx$ 1700 cm$^{-1}$ in both experiment and theory. The agreement is less good at low energy, where the modes are strongly coupled to the environment \cite{Fleury1995}. Altogether, this provides a detailed spectroscopic fingerprint of the single-molecule emitter and its surroundings in a way that ressembles recent tip-enhanced Raman spectroscopy experiments at the single molecule level \cite{Zhang2013}, but without the need of an external optical excitation.

\begin{figure}
  \includegraphics[width=1.00\linewidth]{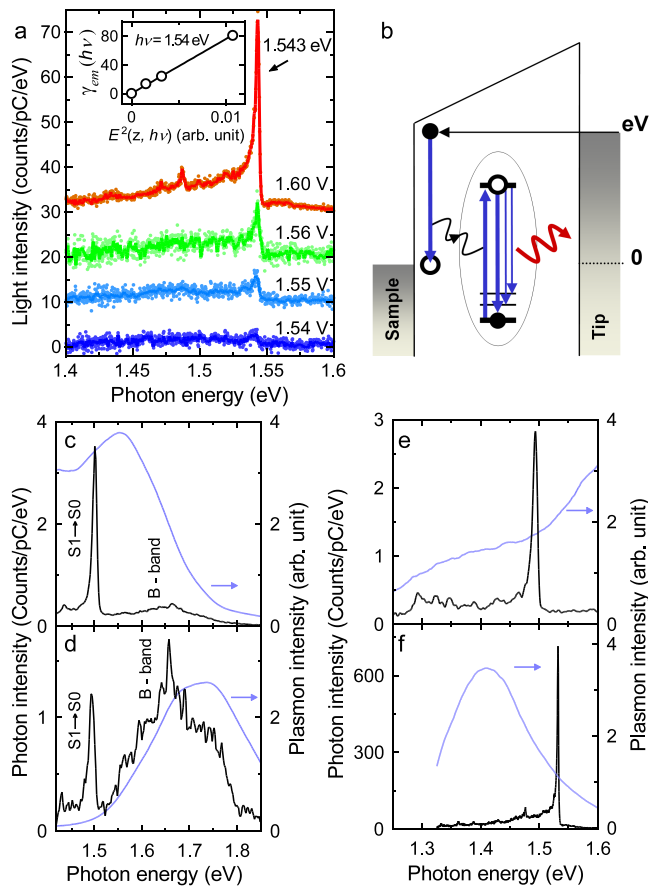}
  \caption{\label{fig4}\textbf{Exciton-plasmon interactions} (a) Light spectra of a suspended copolymer as a function of voltage ($z$ = 2.1 nm, high resolution grating) and (inset) emission rate (in counts/sec) as a function of $E^2(z,h\nu)$ at $h\nu$ = 1.54 eV for the spectra at 1.6 V, 1.56 V and 1.55 V. (b) Sketch of the luminescence mechanism: an inelastic tunnelling electron excites a plasmon that is absorbed by the suspended emitter. (c,d,e,f) Spectra of different copolymer junctions with their respective plasmon spectra (blue line) recorded with the metal tip in front of the bare Au(111) prior to the formation of a molecular junction.} 
\end{figure}

\noindent
We now turn to the exciton-plasmon interactions. The spectra of \Figref{fig4}a show that the voltage onset of the emission matches the energy of the 0-0 peak, a behaviour that is strikingly different from the cases where the emission is interpreted in terms of an intramolecular recombination of electrons and holes injected from the tip and the sample \cite{Qiu2003,Wu2008,Zhu2013,Reecht2014,Lee2014}. In these examples, both S1 and S0 states need to be located between the Fermi levels of the electrodes, a configuration that requires a higher voltage than the energy of the S1$\rightarrow$S0 transition. The behaviour seen in \Figref{fig4}a suggests another mechanism where inelastic tunnelling electrons excite localised plasmons that are finally absorbed by the molecular emitter (\Figref{fig4}b). While similar excitation mechanisms were reported for multilayers of molecules in recent STM experiments \cite{Dong2010,Schneider2012a} and adressed theoretically \cite{Tian2011,Miwa2014}, it has not been identified so far at the level of a single emitter in a STM junction. 

Anger \textit{et al.} \cite{Anger2006} have demonstrated that in this configuration, the emission rate $\gamma_{em}$ of the molecule follows:
\begin{equation}
\gamma_{em} = \gamma_{exc}\;Q ,
\end{equation}
where $\gamma_{exc}$ is the emitter excitation rate and $Q$ its photon emission probability, or quantum yield. The inset of \Figref{fig4}a shows the variation of $\gamma_{em}$ for the 0-0 peak in \Figref{fig4}a as a function of the electromagnetic field intensity at the same energy ($E^2(z, h\nu = 1.54 eV)$) as deduced from metal-metal junction measurements (see supplementary section S6 \cite{supp}). $Q$ is constant for this set of spectra, hence $\gamma_{em}$ only depends on the excitation rate (see supplementary section S6 \cite{supp}). This plot reveals that $\gamma_{exc}$ $\propto$ $E^2(z,h\nu)$, which is the expected behaviour for an optical excitation of the emitter \cite{Anger2006}. This is further illustrated in \Figref{fig4}c and d which show that the intensity ratio between the S1$\rightarrow$S0 and the B-band emission is modified in favour of the transition that experiences the largest plasmon intensity.\\

\noindent
Conversely, \Figref{fig4}e and f show that the relative intensities of the vibronic peaks (which are not affected by variations of the excitation rate) are reduced if the plasmon is intense at the corresponding energies. $Q$ is therefore reduced by the presence of plasmon resonances, a behaviour indicating a high intrinsic quantum yield of the emitter \cite{Anger2006}. Overall, our data show that the surface plasmons enhance the excitation rate of the suspended emitter but reduce its probability to decay radiatively. This also confirms that the highly coherent excited state of our single-molecule junction can be efficiently transferred to the plasmons in a non-radiative decay of the molecular exciton, a necessary condition for future hybrid electronic-plasmonic circuits integrating single molecules as controllable transducer. We believe that the narrow linewidth of this molecular source will enable quantum communication experiments and applications with electrically gated plasmonic circuits.  
\noindent

\noindent
The authors thank Stéphane Berciaud, Fabrice Charra and Laurent Limot for stimulating discussions, and Virginie Speisser, Jean-Georges Faullumel, Michelangelo Romeo and Olivier Cregut for technical support. The Agence National de la Recherche (project SMALL'LED No. ANR-14-CE26-0016-01), the Labex NIE (Contract No. ANR-11-LABX-0058\_NIE), the Région Alsace and the International Center for Frontier Research in Chemistry (FRC) are acknowledged for financial support.

\end{document}